\documentstyle[aps,pra,twocolumn]{revtex}
\begin{document}
\title{$K$-quantum nonlinear coherent states: formulation, realization and
nonclassical effects}
\author{Nguyen Ba An}
\address{National Center for Theoretical Sciences, P. O. Box 2-131, Hsinchu, Taiwan \\
300, R. O. C. }
\maketitle

\begin{abstract}
We introduce a generalized class of states called $K$-quantum nonlinear
coherent states. Each $K$-state has $K$ $j$-components corresponding to one
and the same eigenvalue. Each $Kj$-component can be composed of $K$ $K=1$%
-states in a correlated manner. The introduced states are shown to be
realized in the long-term behavior of the vibrational motion of an ion
properly trapped and laser-driven. Nonclassical properties of the states are
studied in detail.

\end{abstract}


\section{Introduction}

The coherent state (CS) introduced in 1963 \cite{cs1,cs2} has become an
useful and necessary tool to treat ideal boson fields subjected to external
pumping sources. However, CS's cannot describe nonclassical effects such as
antibunching, squeezing, etc. The conventional squeezed states \cite{css}
has been generalized to different types of higher-order ones \cite
{hongmandel,hillery,kumargupta,antinh}, still for ideal bosonic particles.
Elementary excitations in matter, on the other hand, are quasi-particles
often obeying neither Bose-Einstein nor Fermi-Dirac statistics. Recently,
the notion of nonlinear coherent state (NCS) has emerged to adequately deal
with such quasi-particles with commutation relations deformed from the usual
boson/fermion ones. $q$-deformed oscillators \cite{q1,q2} were found as a
particular case of the general $f$-oscillators \cite{f} which possess in
themselves a kinematic nonlinearity causing orbit-dependent oscillation
frequencies. The NCS \cite{ncs1,ncs2,ncs3,ncs4,ncs5} is defined as the
right-hand eigenstate $\left| \xi ;f\right\rangle $ of the nonboson operator 
$A=af(\widehat{n}),$%
\begin{equation}
A\left| \xi ;f\right\rangle =\xi \left| \xi ;f\right\rangle ,  \label{1}
\end{equation}
with $\widehat{n}=a^{+}a,$ $a$ the bosonic annihilation operator, $\xi $ a
complex eigenvalue and $f$ an arbitrary nonlinear operator-valued function
of $\widehat{n}.$ An island where NCS's find their life is single trapped
ions driven by lasers \cite{ncs1,t1,t2,t4,t5}.

In this paper we introduce the $K$-quantum nonlinear coherent state ($K$NCS)
which reduces to the usual NCS defined by Eq. (\ref{1}) when $K=1.$ In
Section II the formulation is given for the $K$NCS together with their
mathematical properties. Section III is devoted to a physical scheme of
generation of $K$NCS's. Their nonclassical properties are studied in detail
in Section IV. Conclusion is the final section.

\section{$K$-quantum nonlinear coherent states}

The $K$NCS is a generalization of the NCS to the $K$ right-hand eigenstates $%
\left| \xi ;Kj,f\right\rangle $ of the non-Hermitian operator $a^{K}f(%
\widehat{n})$ as 
\begin{equation}
a^{K}f(\widehat{n})\left| \xi ;Kj,f\right\rangle =\xi \left| \xi
;Kj,f\right\rangle  \label{2}
\end{equation}
where $K=1,2,...$ and $j=0,1,...,K-1.$ Spanned in the Fock basis $\left|
n\right\rangle ,$%
\begin{equation}
\left| \xi ;Kj,f\right\rangle =\sum_{n=0}^{\infty }c_{n}\left|
n\right\rangle ,  \label{3}
\end{equation}
the expansion coefficients satisfy the recurrence equation 
\begin{equation}
c_{m+K}=\frac{\xi \sqrt{m!}}{\sqrt{(m+K)!}f(m+K)}c_{m}.  \label{4}
\end{equation}
Equation (\ref{4}) determines all the $c$'s if $c_{j}$ with $j=0,1,...,K-1$
are known, 
\begin{equation}
c_{nK+j}=\frac{\xi ^{n}c_{j}}{\sqrt{(nK+j)!}f(nK+j)(!)^{K}},  \label{5}
\end{equation}
where 
\begin{equation}
f(l)(!)^{K}\equiv \left\{ 
\begin{array}{l}
f(l)f(l-K)f(l-2K)...f(m)\text{ if }l\geq K \\ 
1\text{ if }0\leq l\leq K-1
\end{array}
\right. .  \label{6}
\end{equation}
Of course, $0\leq m\leq K-1,$ in Eq. (\ref{6}). The $c_{j}$ in Eq. (\ref{5})
are themselves determined by the normalization condition $\left\langle \xi
;Kj,f\right| \left. \xi ;Kj,f\right\rangle =1$ which yields 
\begin{eqnarray}
c_{j} &\rightarrow &c_{Kj}\equiv c_{Kj}(|\xi |^{2})  \nonumber \\
&=&\left[ \sum_{m=0}^{\infty }\frac{|\xi |^{2m}}{(mK+j)!\left|
f(mK+j)(!)^{K}\right| ^{2}}\right] ^{-1/2}.  \label{7}
\end{eqnarray}
In principle, the $c_{Kj}$ is uncertain up to a phase factor. In Eq. (\ref{7}%
) we have chosen the phase such that the usual CS results in the limit $%
f\equiv 1$ and $K=1.$ The explicit form of the $K$NCS is then 
\begin{equation}
\left| \xi ;Kj,f\right\rangle =c_{Kj}\sum_{n=0}^{\infty }\frac{\xi ^{n}}{%
\sqrt{(nK+j)!}f(nK+j)(!)^{K}}\left| nK+j\right\rangle .  \label{8}
\end{equation}
Unlike the usual CS whose eigenvalues spread the whole complex plane, those
of the $K$NCS are bounded in each particular case. This limitation is felt
from Eq. (\ref{8}) that requires $c_{Kj}$ not be zeros. Yet, the $c_{Kj}$
given by Eq. (\ref{7}) would vanish if the sum over $m$ diverges. The
constraint for the existence of $K$NCS's, i.e. for $c_{Kj}$ not to be zeros,
is thus imposed simultaneously on $\xi ,$ $f$ and $K$ so that 
\begin{equation}
\lim_{m\rightarrow \infty }\left\{ |\xi |^{2m}\left[
(mK+j)!|f(mK+j)(!)^{K}|^{2}\right] ^{-1}\right\} =0.  \label{9}
\end{equation}
The importance of this constraint will be seen in the next section when the
specific nonlinear function $f$ is addressed.

For a given $K$ the eigenvalue $\xi $ is $K$-degenerate. The $K$
corresponding eigenfunctions $\left| \xi ;Kj,f\right\rangle $ with $%
j=0,1,...,K-1$ are orthogonal to each other, 
\begin{equation}
\left\langle \xi ;Kj,f\right| \left. \xi ;Kj^{\prime },f\right\rangle
=\delta _{jj^{\prime }},  \label{10}
\end{equation}
as easily verified from the expansion (\ref{8}). However, with distinct
eigenvalues $\xi $ and $\xi ^{\prime }\neq \xi ,$%
\begin{equation}
\left\langle \xi ;Kj,f\right| \left. \xi ^{\prime };Kj,f\right\rangle =\frac{%
c_{Kj}(|\xi ^{\prime }|^{2})c_{Kj}(|\xi |^{2})}{c_{Kj}^{2}((\xi ^{\prime
})^{*}\xi )}\neq 0.  \label{11}
\end{equation}
Because of the non-orthogonality (\ref{11}), the $K$NCS's constitute an
overcomplete set with the following resolution of unity 
\begin{equation}
\int \frac{d^{2}\xi }{\pi }\sum_{j=0}^{K-1}\left| \xi ;Kj,f\right\rangle
\left\langle \xi ;Kj,f\right| =1.  \label{12}
\end{equation}

This section is ended by an interesting observation that any state $\left|
\xi ;Kj,f\right\rangle $ can be decomposed into a linear superposition of $K$
states $\left| \xi _{j^{\prime }};10,f\right\rangle .$ Namely, 
\begin{equation}
\left| \xi ;Kj,f\right\rangle =\sum_{j^{\prime }=0}^{K-1}\zeta _{jj^{\prime
}}\left| \xi _{j^{\prime }};10,f\right\rangle ,  \label{13}
\end{equation}
with 
\begin{equation}
\xi _{j^{\prime }}=\xi ^{1/K}\exp \left( \frac{2\pi ij^{\prime }}{K}\right)
\label{14}
\end{equation}
and 
\begin{equation}
\zeta _{jj^{\prime }}=\frac{\xi ^{-j/K}}{K}\frac{c_{Kj}(|\xi |^{2})}{%
c_{10}(|\xi |^{2/K})}\exp \left( -\frac{2\pi ijj^{\prime }}{K}\right) .
\label{15}
\end{equation}
The verification is straightforward by substituting Eqs. (\ref{8}), (\ref{14}%
) and (\ref{15}) into the r.h.s. of Eq. (\ref{13}) with subsequent use of
the identities 
\begin{equation}
\sum_{q=0}^{P-1}\exp \left[ \frac{2\pi i}{P}(l-l^{\prime })q\right] =P\delta
_{ll^{\prime }}
\end{equation}
and 
\begin{equation}
\sum_{n=0}^{\infty }b_{n}\left| n\right\rangle \equiv
\sum_{q=0}^{P-1}\sum_{m=0}^{\infty }b_{mP+q}\left| mP+q\right\rangle
\end{equation}
for arbitrary coefficients $b_{n}$ and integer $P.$

\section{Physical realization}

Motivated by the physical scheme proposed in \cite{ncs1} we consider a
single two-level ion trapped by a harmonic potential $V=kx^{2}/2$ with $k$
the trapping force and $x$ the ion's center-of-mass position. In the
quantized regime the ion vibrates around $x=0$ with frequency $\nu =\sqrt{k/M%
},$ $M$ the ion mass. As a whole, the ion Hamiltonian is ($\hbar =c=1$
throughout) 
\begin{equation}
H_{0}=\Delta \sigma _{3}+\nu a^{+}a  \label{16}
\end{equation}
where $\Delta $ is the energy gap between the two electronic levels of the
ion, $a$ is the bosonic annihilation operator of a quantum of the quantized
vibration of the ion and, $\sigma _{3}$ together with $\sigma _{\pm }$ are
the pseudospin-$\frac{1}{2}$ operators satisfying the commutation relations 
\begin{equation}
\left[ \sigma _{3},\sigma _{\pm }\right] =\pm \sigma _{\pm },\text{ }\left[
\sigma _{+},\sigma _{-}\right] =2\sigma _{3}.  \label{17}
\end{equation}
By controlling the trapping potential the vibration frequency $\nu $ can be
made large enough for the sidebands due to the ion quantized motion to be
well-resolved. Next, the trapped ion is manipulated by laser beams which
couple the ion's electronic to its vibrational degrees of freedom via the
light-matter interaction Hamiltonian 
\begin{equation}
H_{int}=\sum_{l=1}^{L}\Omega _{l}\exp \left( \omega _{l}t+\varphi
_{l}\right) g_{l}(\widehat{x})\sigma _{-}+\text{ h.c.}  \label{18}
\end{equation}
In Eq. (\ref{18}) $L$ is the number of driving lasers, $\Omega _{l}$ the
pure electronic transition Rabi frequencies, $\omega _{l}$ ($\varphi _{l}$)
the laser frequencies (phases) and $g_{l}(\widehat{x})$ the laser spatial
profile acting on the ion through its position operator 
\begin{equation}
\widehat{x}=\frac{1}{\sqrt{2M\nu }}\left( a^{+}+a\right) .  \label{19}
\end{equation}
For traveling waves 
\begin{equation}
g_{l}(\widehat{x})=\exp \left( \frac{-2\pi i\widehat{x}}{\lambda _{l}}%
\right) =\exp \left[ -i\eta \left( a^{+}+a\right) \right]  \label{20}
\end{equation}
with $\lambda _{l}$ the laser wavelengths and $\eta \simeq \eta _{l}=2\pi
/(\lambda _{l}\sqrt{2M\nu }),$ the Lamb-Dicke parameter. In the resolved
sideband limit the lasers can be tuned so as 
\begin{equation}
\omega _{l}=\Delta +n_{l}\nu  \label{21}
\end{equation}
where $n_{l}=1,2,...$ ($n_{l}=-1,-2,...$) imply blue (red) detuning while
resonant tuning corresponds to $n_{l}=0.$ In the interaction representation
associated with ${\cal H}_{0}=H_{0}$ the laser-driven trapped ion is
described by the total Hamiltonian 
\begin{equation}
{\cal H}={\cal H}_{0}+{\cal H}_{int}
\end{equation}
with 
\begin{equation}
{\cal H}_{int}=\exp \left( iH_{0}t\right) H_{int}\exp (-iH_{0}t).  \label{22}
\end{equation}
Putting Eqs. (\ref{16}) and (\ref{18}) into the r.h.s. of Eq. (\ref{22})
with use of the Baker-Hausdorff identity and the relations (\ref{17}) yields 
\begin{equation}
{\cal H}_{int}=F\sigma _{-}+\sigma _{+}F^{+}  \label{23}
\end{equation}
with $F$ given by 
\begin{eqnarray}
F &\equiv &F\left( \eta ,\widehat{n},a^{+},a\right) =  \nonumber \\
&&\exp (-\frac{\eta ^{2}}{2})\sum_{l=1}^{L}\left\{ (-i\eta )^{|n_{l}|}\Omega
_{l}\text{ }\left( a^{+}\right) ^{(|n_{l}|-n_{l})/2}\right.  \nonumber \\
&&\left. \left[ \sum_{m=0}^{\infty }\frac{(-\eta ^{2})^{m}\widehat{n}!}{%
(m+|n_{l}|)!m!(\widehat{n}-m)!}\right] a^{(|n_{l}|+n_{l})/2}\right\} .
\label{24}
\end{eqnarray}
In deriving Eq. (\ref{24}) terms oscillating with frequencies $m\nu $ ($%
m\neq 0$) were omitted since those average to zero in the resolved sideband
limit we are interested in.

The system state $\Psi (t)$ evolves in time following the equation 
\begin{equation}
i\frac{d\Psi }{dt}={\cal H}_{int}\Psi .  \label{25}
\end{equation}
We look for long-term nontrivial steady states $\Psi _{S}$ in which the
internal and external degrees of freedom of the ion become decoupled, i.e. 
\begin{equation}
{\cal H}_{int}\Psi _{S}=0.  \label{26}
\end{equation}
Such nontrivial solutions would be either $\left| \uparrow \right\rangle
\left| \xi \right\rangle $ or $\left| \downarrow \right\rangle \left| \xi
\right\rangle $ where $\left| \uparrow \right\rangle $ ($\left| \downarrow
\right\rangle $) is the electronic excited (ground) state and $\left| \xi
\right\rangle $ describes the vibrational state. The state $\left| \uparrow
\right\rangle \left| \xi \right\rangle $ is unstable due to spontaneous
emission which was not treated explicitly here. The remaining ``dark''state $%
\left| \downarrow \right\rangle \left| \xi \right\rangle $ is stable with $%
\left| \xi \right\rangle $ satisfying the condition 
\begin{equation}
F^{+}\left| \xi \right\rangle =0.  \label{27}
\end{equation}
Similar ``dark'' states have been obtained as an ansatz of master equations
with spontaneous emission included \cite{t1,ncs1}. From Eqs. (\ref{25}) and (%
\ref{27}) it follows that the nontrivial state $\left| \xi \right\rangle $
exists iff there are at least two driving lasers. For generality, let there
be $P$ ($Q$) red-detuned (blue-detuned) laser beams and a single resonant
one. Then, the coefficients of the expansion of $\left| \xi \right\rangle $
in the Fock space, 
\begin{equation}
\left| \xi \right\rangle =\sum_{l=0}^{\infty }C_{l}\left| l\right\rangle ,
\label{28}
\end{equation}
are recurrent as 
\[
\sum_{p=1}^{P}\left( i\eta \right) ^{|n_{p}|}\Omega _{p}\text{e}^{-i\varphi
_{p}}\sqrt{\frac{l!}{(l+|n_{p}|)!}}L_{l}^{|n_{p}|}(\eta ^{2})C_{l+|n_{p}|} 
\]
\[
+\sum_{q=1}^{Q}\theta \left( l-n_{q}\right) \left( i\eta \right)
^{n_{q}}\Omega _{q}\text{e}^{-i\varphi _{q}}\sqrt{\frac{(l-n_{q})!}{l!}}%
L_{l-n_{q}}^{n_{q}}(\eta ^{2})C_{l-n_{q}} 
\]
\begin{equation}
+\Omega _{0}\text{e}^{-i\varphi _{0}}L_{l}^{0}(\eta ^{2})C_{l}=0  \label{29}
\end{equation}
where $\theta \left( x\right) $ is the step function and $L_{n}^{m}(x)$ the $%
n$-th generalized Laguerre polynomial in $x$ for parameter $m$.

For the purpose of generating the $K$NCS defined in the preceding section we
restrict to $P=1,$ $n_{p=1}=-K$ and $Q=0$ and get from Eq. (\ref{29}) 
\begin{equation}
C_{l+K}=-\frac{\text{e}^{i\varphi }\Omega _{0}\sqrt{(l+K)!}L_{l}^{0}(\eta
^{2})}{(i\eta )^{K}\Omega _{1}\sqrt{l!}L_{l}^{K}(\eta ^{2})}C_{l}  \label{30}
\end{equation}
with $\varphi =\varphi _{1}-\varphi _{0}.$ Comparing Eqs. (\ref{30}) and (%
\ref{4}) indicates that $\left| \xi \right\rangle $ obeying Eq. (\ref{27})
is a $K$NCS with the physical controllable eigenvalue 
\begin{equation}
\xi =-\frac{\text{e}^{i\varphi }\Omega _{0}}{(i\eta )^{K}\Omega _{1}}
\label{31}
\end{equation}
and the specific, also controllable, nonlinear function 
\begin{equation}
f\left( \widehat{n}+K\right) =\frac{\widehat{n}!L_{\widehat{n}}^{K}(\eta
^{2})}{\left( \widehat{n}+K\right) !L_{\widehat{n}}^{0}(\eta ^{2})}.
\label{32}
\end{equation}

In the next section the number distribution, squeezing and antibunching of
the $K$NCS with the specific $\xi $ and $f$ given above will be studied in
detail. To that aim let us define the function 
\begin{equation}
g_{Kj}(n,l)=\frac{\xi ^{n+l/K}}{\sqrt{(nK+j+l)!}\text{ }f(nK+j+l)(!)^{K}}
\label{33}
\end{equation}
in terms of which 
\begin{equation}
\left| \xi ;Kj,f\right\rangle =\frac{\sum_{n=0}^{\infty }g_{Kj}(n,0)\left|
nK+j\right\rangle }{\sqrt{\sum_{m=0}^{\infty }\left| g_{Kj}(m,0)\right| ^{2}}%
}.  \label{34}
\end{equation}
It is worth to emphasize that the $K$NCS does not exist for arbitrary
control parameters $\eta $ and $\xi .$ For a fixed trapping potential
characterized by $\eta $ the amplitudes (phases are not sensitive in this
scheme with $Q=0$ and $P=1$) of the driving fields must be chosen such that
the constraint (\ref{9}) is met, i.e. $\lim_{n\rightarrow \infty
}|g_{Kj}(n,0)|^{2}=0.$ Figure 1, log($|g_{Kj}(n,0)|^{2}$) versus $n,$ shows
opposite limiting behaviors for the same value of $\eta $ but different $%
|\xi |.$ In general, there is a critical $|\xi _{c}|$ above which the $K$NCS
vanishes. This critical $|\xi _{c}|$ depends strongly on both $\eta $ and $K$
but not on $j.$ Phase diagrams for the existence domain of the $K$NCS are
drawn in Fig. 2 in the ($\eta $,$|\xi |$)-plane for $K=1,2$ and $3.$ These
diagrams guide the appropriate experimental choice of the control parameters
to observe the $K$NCS with a concrete $K.$

\section{Nonclassical effects}

\subsection{Multi-peaked number distribution}

The probability of finding $n$ quanta in the $K$NCS, i.e. its number
distribution, is given by 
\begin{equation}
P_{Kj}(n)=\frac{I(\frac{n-j}{K})}{\sum_{m=0}^{\infty }\left|
g_{Kj}(m,0)\right| ^{2}}\left| g_{Kj}(\frac{n-j}{K},0)\right| ^{2}
\label{35}
\end{equation}
where $I(x)$ equals $x$ if $x$ is a non-negative integer and zero otherwise.
The tortuous shape in Fig. 1 gives rise to a multi-peaked structure of $%
P_{Kj}(n)$ as depicted in Fig. 3 in the case $K=1,$ $j=0.$ The multi-peaked
distribution is very peculiar compared to the Poisson's in the usual CS.
However, multiple peaking disappears in the Lamb-Dicke limit $\eta \ll 1$ in
which $\left| g_{Kj}(n,0)\right| $ varies monotonically (not tortuously) for
increasing $n$ having only one peak at some value $n>0$ and then decreasing
quickly. The developing from single- to multi-peaked structure as $\eta $
increases can be called self-splitting \cite{ncs1} which would bring about
abundant nonclassical phenomena including quantum interferences if the peaks
are located nearby and comparable in heights. As for higher orders $K,$
unexpectedly, the self-splitting tends to be less pronounced as seen from
Fig. 4. Besides getting less tortuous, the more important fact is that the
curve decreases much faster for higher $K.$ Hence, for $K>1$ the $\eta $%
-governed self-splitting is negligible physically. Instead, for $K>1,$
another kind of splitting, the $K$-governed one, appears independent of the
Lamb-Dicke parameter. This is dictated by the presence of $I(\frac{n-j}{K})$
in Eq. (\ref{35}). Namely, for $K>1$ the number distribution ``oscillates''
with ``period'' $K$ and its ``phase'' is $j$-dependent. The $P_{30},$ $%
P_{31} $ and $P_{32}$ plotted in Fig. 5 illustrate that, in the state $%
\left| \xi ;30,f\right\rangle $ ($\left| \xi ;31,f\right\rangle ;$ $\left|
\xi ;32,f\right\rangle $), finite is only the probability of finding $%
3,6,9,...$ ($4,7,10,...$; $5,8,11,...$) quanta. Generally speaking, only a
number of quanta equal to a multiple $K$ plus $j$ ($j=0,$ $1,...,$ $K-1$)
can be found in the state $\left| \xi ;Kj,f\right\rangle .$ Given $K$ each $%
j $-state occupies its own subspace and all the different $j$-states fill
the entire Fock space. Therefore, the notion ``number-parity'' might be
introduced. A state is said to have a number-parity $Kj$ if it contains only 
$nK+j$ ($n=0,$ $1,$ $2,...$) quanta. For $K=2$ the parity gets its intuitive
meaning: the state $\left| \xi ;20,f\right\rangle $ ($\left| \xi
;21,f\right\rangle $) contains only even (odd) numbers of quanta and it can
be referred to as even (odd) NCS \cite{eo1,eo2}. This evenness (oddness) is
also clear from the view point of the decomposition (\ref{13}) according to
which 
\begin{equation}
\left| \xi ;20,f\right\rangle =\zeta _{00}\left| \xi _{0};10,f\right\rangle
+\zeta _{01}\left| \xi _{1};10,f\right\rangle ,  \label{e}
\end{equation}
\begin{equation}
\left| \xi ;21,f\right\rangle =\zeta _{10}\left| \xi _{0};10,f\right\rangle
+\zeta _{11}\left| \xi _{1};10,f\right\rangle .  \label{o}
\end{equation}
By virtue of Eqs. (\ref{14}) and (\ref{15}), $\xi _{0}=-\xi _{1}$ and $\zeta
_{00}=\zeta _{01}=\zeta _{10}=-\zeta _{11}.$ These mean that the combination
in Eq. (\ref{e}) is symmetric leading to evenness, while that in Eq. (\ref{o}%
) is anti-symmetric leading to oddness.

Because the ${\cal H}_{int}$ determined in Eqs. (\ref{23}) and (\ref{24})
for $P=1,$ $n_{p=1}=-K,$ $Q=0$ creates/annihilates $K$ quanta at a time the
number-parity is conserved. If initially the vibration quanta are in a Fock
state $\left| m\right\rangle $ then the steady state $\Psi _{S}$ will have
number-parity $Kj$ with $j$ the remainder of $m$ divided by $K.$ By sideband
cooling modern techniques the fundamental limit has been reached in which
the ion vibrational motion can be kept to zero-point energy 98\% (92\%) of
the time in 1D (3D) \cite{zero}. If such a ground state is further
manipulated by lasers as proposed above, then the state $\left| \xi
;K0,f\right\rangle $ will be realized. If, the initial quanta are prepared
in a CS $\left| \alpha \right\rangle $ then the eventual state will be a
mixed state with weighted $j$-components 
\begin{equation}
\Psi _{S}=\Psi _{K}^{mix}=\sum_{j=0}^{K-1}\beta _{Kj}\left| \xi
;Kj,f\right\rangle .  \label{36}
\end{equation}
Assuming the weights $\beta _{Kj}$ be as in the CS, 
\begin{equation}
\beta _{Kj}=\exp \left( -\frac{|\alpha |^{2}}{2}\right) \sqrt{%
\sum_{m=0}^{\infty }\frac{|\alpha |^{2(mK+j)}}{(mK+j)!}},  \label{37}
\end{equation}
the number distribution $P_{K}^{mix}$ of the mixed state $\Psi _{K}^{mix}$
will differ noticeably under two situations: $|\alpha |<K$ and $|\alpha
|\geq K.$ For $|\alpha |<K$ the $\beta _{Kj}$ are scattered whereas for $%
|\alpha |\geq K$ they get equal, $\beta _{Kj}=1/\sqrt{K},$ for all $j$
reducing $P_{K}^{mix}$ simply to 
\begin{equation}
P_{K}^{mix}(n)=\frac{1}{K}\left( \sum_{j=0}^{K-1}\sqrt{P_{Kj}(n)}\right) ^{2}
\label{38}
\end{equation}
with $P_{Kj}(n)$ given in Eq. (\ref{35}), independent of $\alpha .$ Figure 6
addresses the above-said issue for $K=6$ and several values of $|\alpha |.$
The $P_{K}^{mix}(n)$ spans the whole Fock space with different $\alpha $%
-dependent profiles. This is of course subject to experimental checking to
which extent the assumption (\ref{37}) is valid.

\subsection{Squeezing and antibunching}

In the small $\eta $ limit the $K=1$ NCS exhibits a single-peaked number
distribution that may be super- or sub-poissonian. So happens as well for
the envelope of the $K>1$ NCS distribution. Sub-poisson distribution
featuring antibunching occurs whenever the Mandel parameter, 
\begin{equation}
M=\frac{\left\langle \widehat{n}^{2}\right\rangle }{\left\langle \widehat{n}%
\right\rangle }-\left\langle \widehat{n}\right\rangle ,  \label{39}
\end{equation}
is less than $1$. We shall also examine amplitude-quadrature squeezing of
the ion center-of-mass position operator (in units of $(2M\nu )^{-1/2}$, for
convenience) $\overline{x}=a^{+}+a$ by calculating its variance $%
\left\langle \left( \Delta \overline{x}\right) ^{2}\right\rangle =1+2S$ with 
$S$ given by 
\begin{equation}
S=\left\langle \widehat{n}\right\rangle +\Re \left\langle a^{2}\right\rangle
-2\Re ^{2}\left\langle a\right\rangle .  \label{40}
\end{equation}
The $K$NCS gets squeezed in $\overline{x}$ if $-0.5\leq S<0.$ In terms of
the functions $g_{Kj}(m,l)$ defined in Eq. (\ref{33}) 
\begin{equation}
\left\langle a^{l}\right\rangle _{Kj}=\frac{\sum_{m=0}^{\infty }\left[ \sqrt{%
\frac{(mK+j+l)!}{(mK+j)!}}g_{Kj}^{*}(m,0)g_{Kj}(m,l)\right] }{%
\sum_{m=0}^{\infty }\left| g_{Kj}(m,0)\right| ^{2}},  \label{41}
\end{equation}
\begin{equation}
\left\langle \widehat{n}^{l}\right\rangle _{Kj}=\frac{\sum_{m=0}^{\infty
}\left[ \left( mK+j\right) ^{l}\left| g_{Kj}(m,0)\right| ^{2}\right] }{%
\sum_{m=0}^{\infty }\left| g_{Kj}(m,0)\right| ^{2}}.  \label{42}
\end{equation}
The analytic expressions of $M$ and $S$ are readily obtained using Eqs. (\ref
{41}) and (\ref{42}) in (\ref{39}) and (\ref{40}). For $K=1$ it is found
that $S>0,$ i.e. no squeezing, but $M$ decreases from $1$ for increasing $%
|\xi |$ (see Fig. 7), i.e. antibunching. Such effects are opposed to the
usual CS for which $S=0$ and $M=1$ for all $|\xi |.$ As was well-known, for $%
f\equiv 1$ the state $\left| \xi ;20,1\right\rangle $ possesses squeezing
but no antibunching whereas the state $\left| \xi ;21,1\right\rangle $ does
the inverse: antibunching but no squeezing. The $K$NCS with the specific
nonlinear function $f$ given by Eq. (\ref{32}) causes curious changes. The
state $\left| \xi ;20,f\right\rangle $ is squeezed in a rather narrow range
of small values of $|\xi |$ (see Fig. 8a) and turns from super- to
sub-poisson's as $|\xi |$ increases (see Fig. 8b). As for the state $\left|
\xi ;21,f\right\rangle $ it remains unsqueezed but antibunched for all $|\xi
|$ as were no affect of $f.$ For $K=3$, the state $\left| \xi
;30,f\right\rangle $ behaves qualitatively like the state $\left| \xi
;20,f\right\rangle $ does. The state $\left| \xi ;31,f\right\rangle $ is
non-squeezed but, most curiously, its number distribution crosses $1$ twice
(see Fig. 9). At small values of $|\xi |$, the Mandel parameter $M_{31}$ is
less than $1.$ It crosses $1$ becoming super-poissonian but, after reaching
a maximum, drops back below $1$ and remains there being sub-poissonian in
the whole high-value side of $|\xi |.$ Finally, no squeezing and permanent
antibunching are found in the state $\left| \xi ;32,f\right\rangle .$

\section{Conclusion}

We have introduced the $K$-quantum nonlinear coherent state. For a given $K$
there are $K$ mutually orthogonal substates belonging to the same
eigenvalue. Each of such substates can be decomposed into a linear
combination of $K$ usual nonlinear coherent states with correlated
eigenvalues and weights determined respectively by Eqs. (\ref{14}) and (\ref
{15}). States with the same $K$ and $j$ but belonging to two different
eigenvalues $\xi $ and $\xi ^{\prime }\neq \xi $ are however non-orthogonal
to each other and, the states with a fixed $K$ and all possible $\xi $ and $%
j $ are overcomplete. A harmonically trapped two-level ion driven properly
by two laser beams, one in resonance with the ion's electronic transition
and the other detuned to the $K^{th}$ lower sideband, is shown to realize
the $K$-quantum nonlinear coherent state in the steady regime for suitably
chosen control parameters. These states show up two kinds of multiple
self-peaking in their number distribution. One kind is associated with large
values of the Lamb-Dicke parameter $\eta $ but is not pronounced for $K>1.$
The other kind originates from $K>1$ independent of $\eta $ and is
manifested in the number distribution that fills the Fock space
``periodically'' with ``period'' $K,$ forming a multi-peaked structure.
Squeezing and antibunching have been investigated in detail revealing
curious features corresponding to the specific nonlinear function $f(%
\widehat{n})$ which is here well-determined in the context of a trapped ion
driven by lasers with the certain setup. Another setup of driving lasers,
say, with $P=0$ and $Q=1$, or whatever else, would lead to another class of
states.

\begin{center}
{\bf ACKNOWLEDGMENTS}
\end{center}

This work was supported by the National Center for Theoretical Sciences,
Physics Division, Hsinchu, Taiwan, R.O.C.

\newpage

\begin{center}
{\Large FIGURE CAPTIONS}
\end{center}

\begin{description}
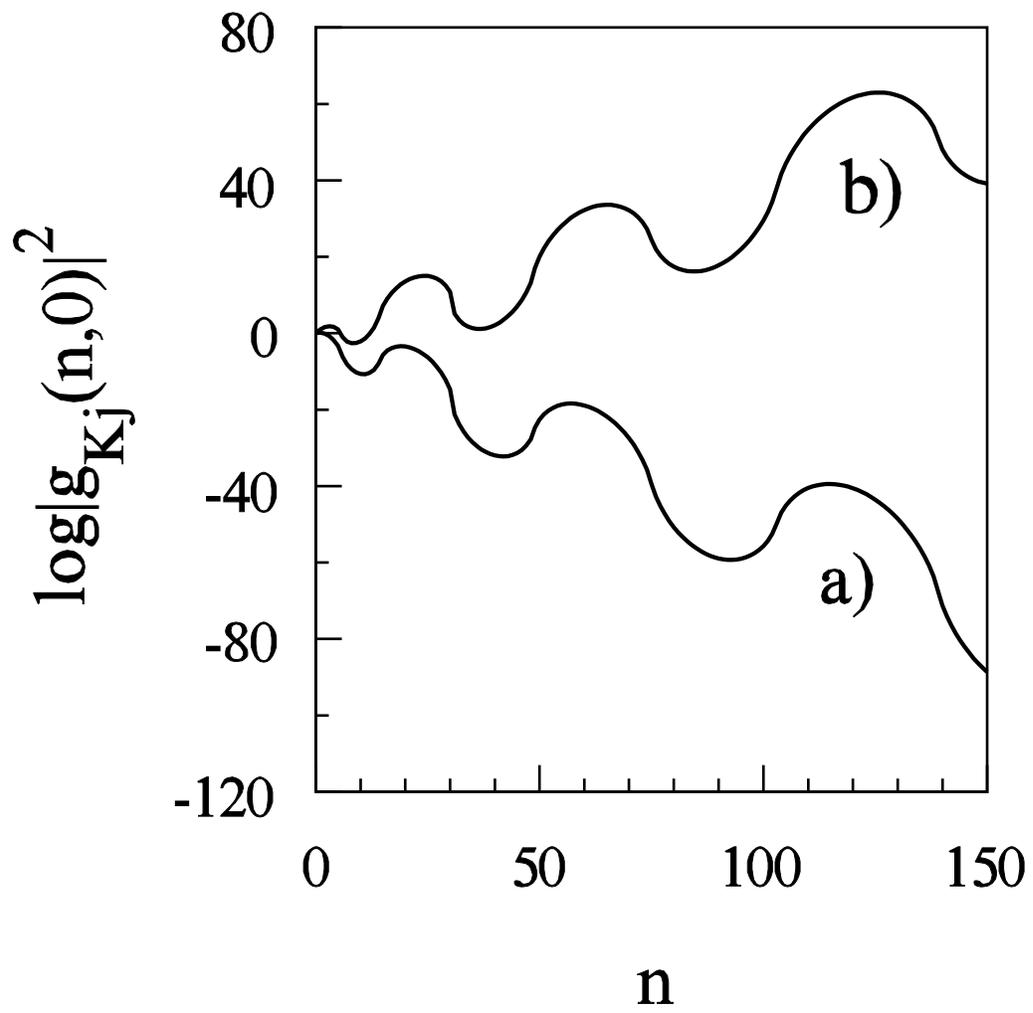
\item[Fig. 1:]  log($|g_{Kj}(n,0)|^{2}$) as a function of $n$ for $K=1,$ $%
j=0,$ $\eta =0.5$ while a) $|\xi |=1.2,$ a converging case, and b) $|\xi
|=3.2,$ a diverging case.

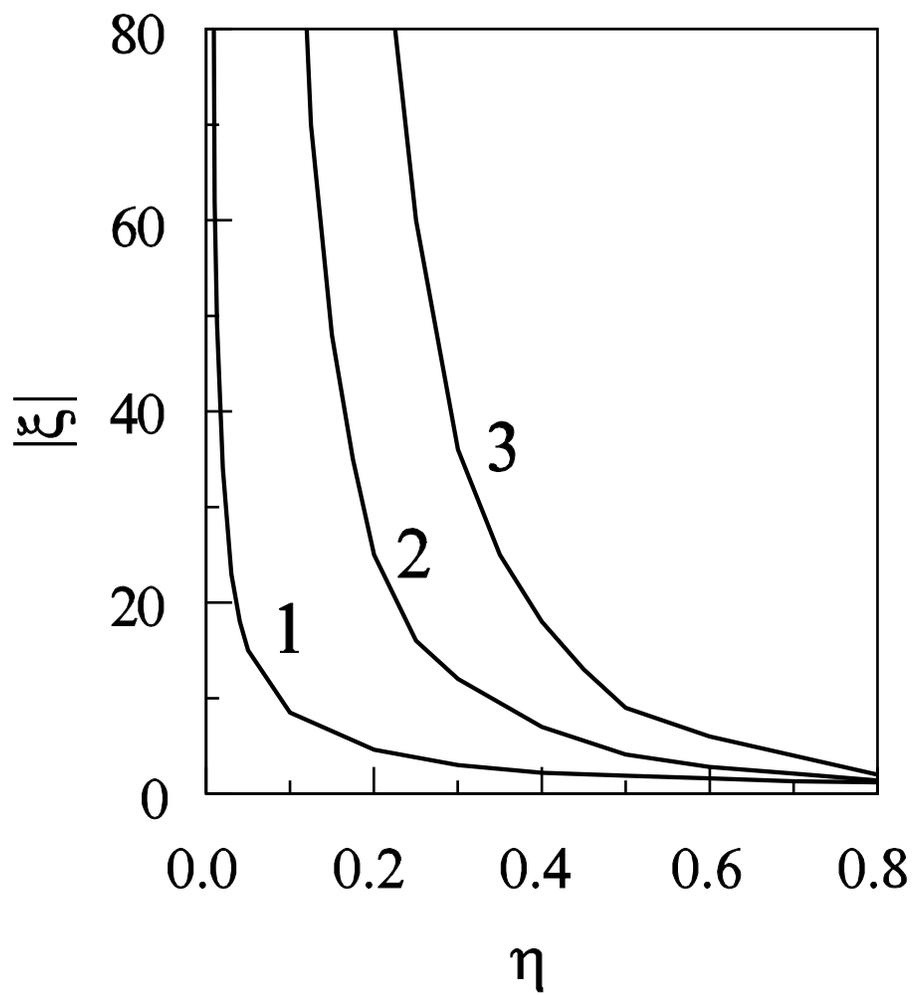
\item[Fig. 2:]  Phase diagrams in the ($\eta $,$|\xi |$)-plane for various $K
$ whose values are indicated near the curve. The curves, themselves, are the 
$|\xi _{c}|$ as a function of $\eta .$ The corresponding $K$NCS exists (does
not exist) below (above) the curve.

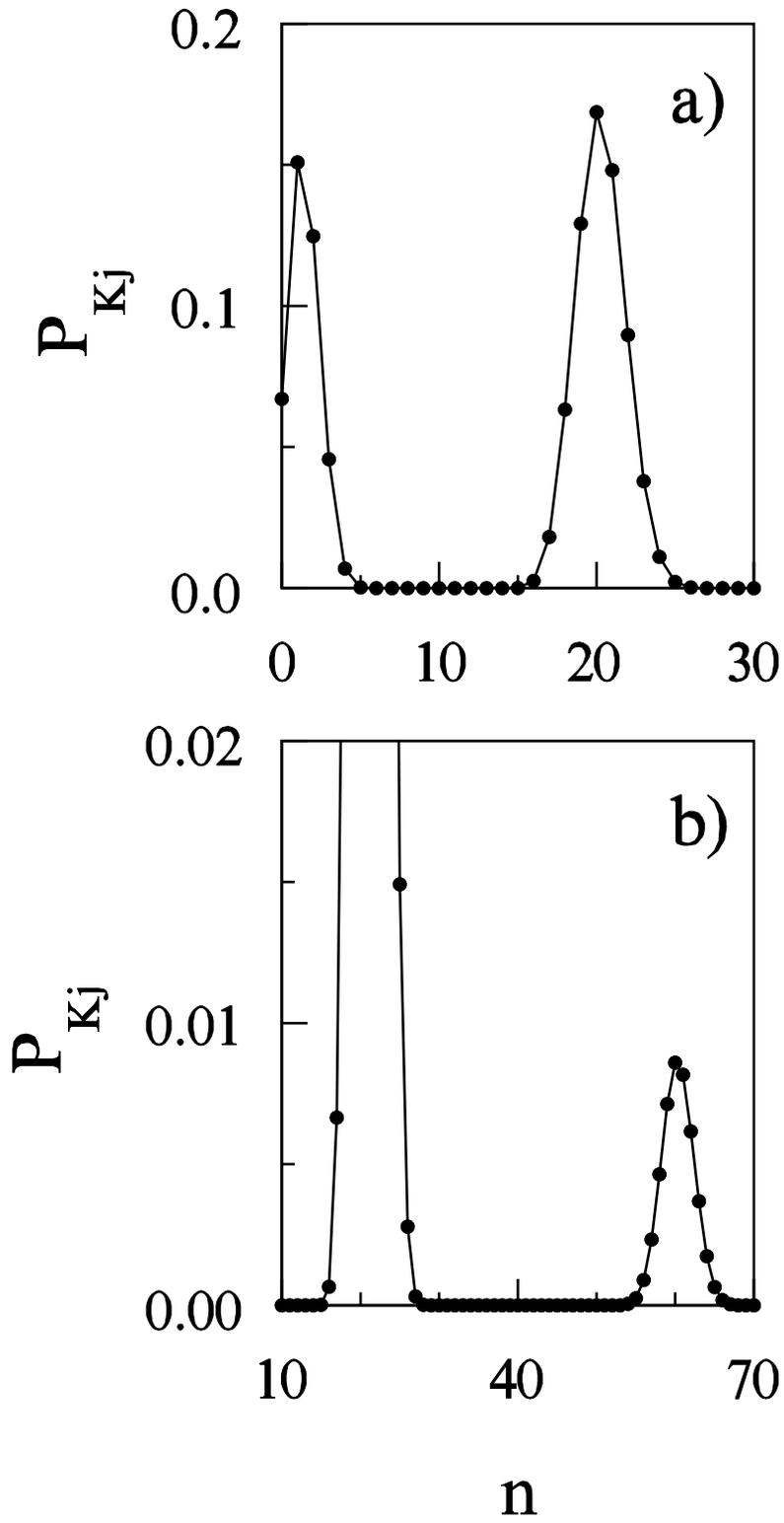
\item[Fig. 3:]  Number distributions for the states $\left| \xi
;10,f\right\rangle $ with $\eta =0.5$ while a) $|\xi |=1.5$ and b) $|\xi
|=1.8,$ featuring two-peaked structures.

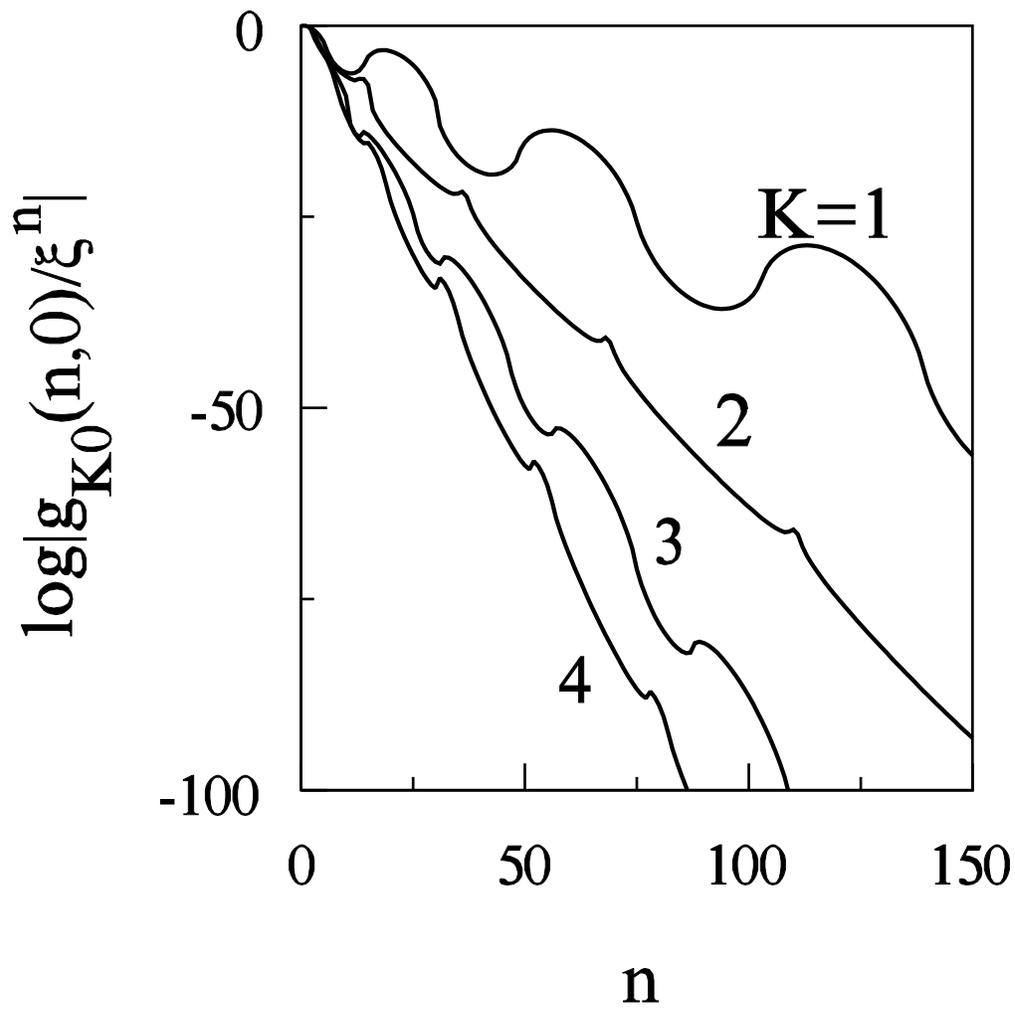
\item[Fig. 4:]  log($|g_{K0}(n,0)/\xi ^{n}|$) versus $n$ for $\eta =0.5$ and
different orders $K$ indicated near the curve. The curve gets less tortuous
and drops much faster for a higher $K.$

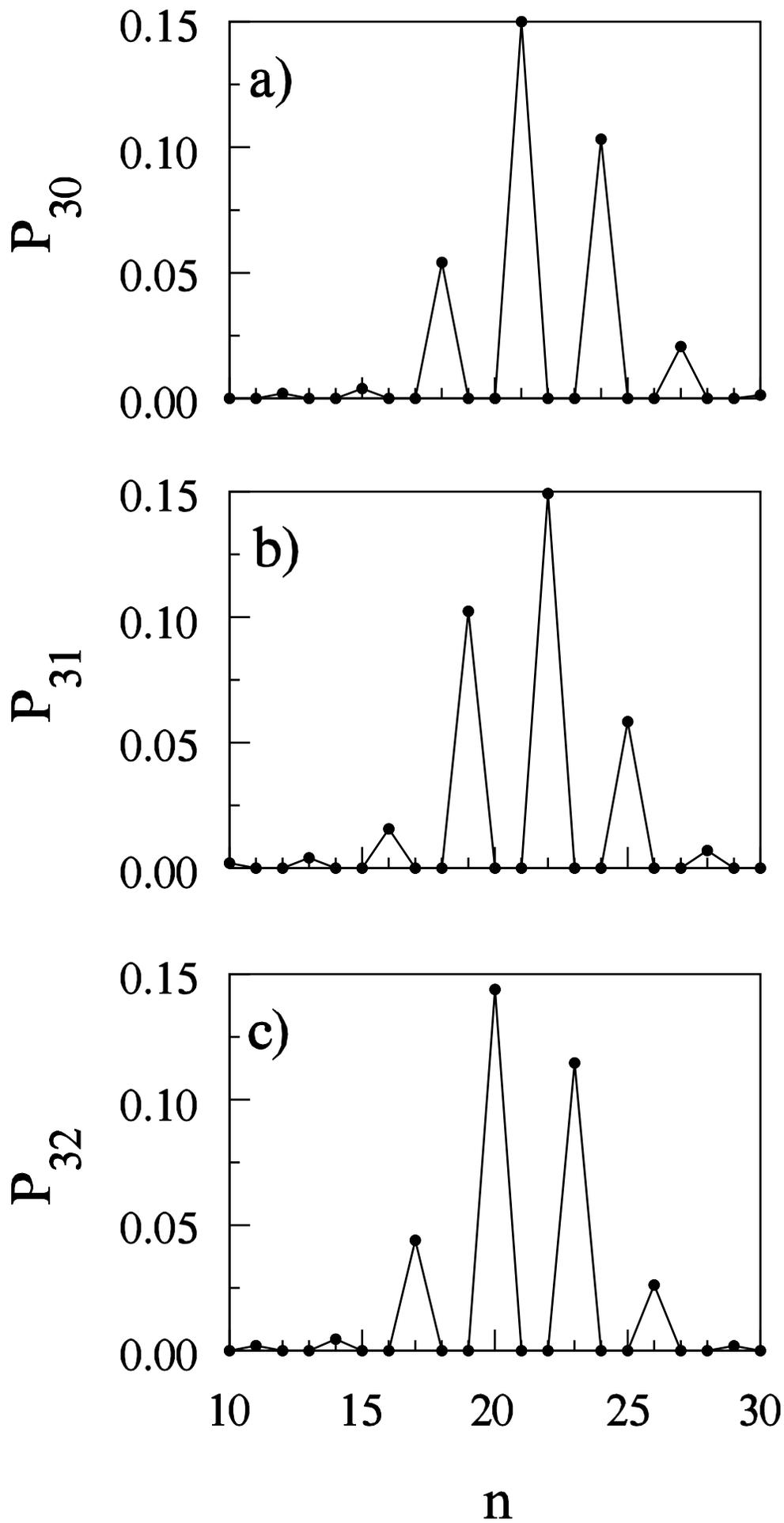
\item[Fig. 5:]  Number distributions $P_{Kj}$ versus $n$ for $\eta =0.05,$ $%
|\xi |=100,$ $K=3$ while a) $j=0$, b) $j=1$ and c) $j=2,$ featuring
multi-peaked structures.

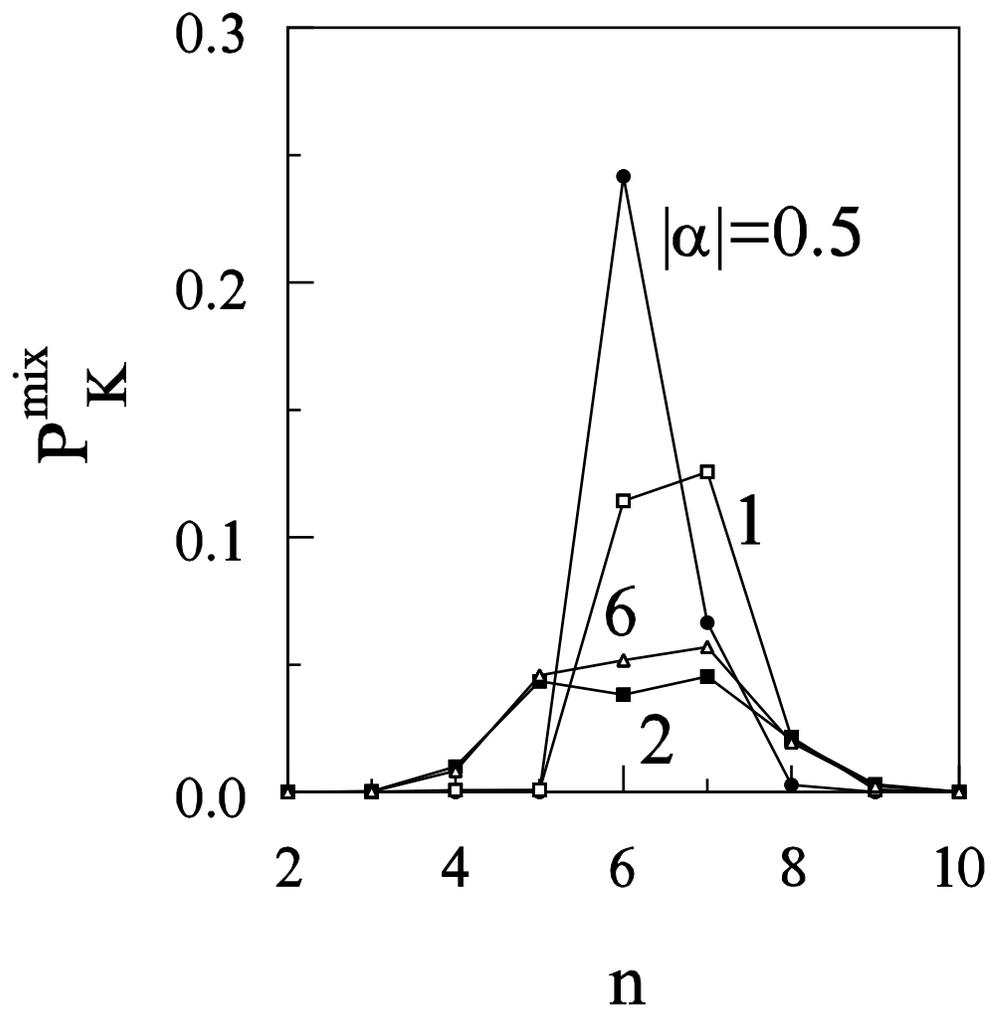
\item[Fig. 6:]  Number distributions of the mixed state resulting from an
initial coherent state $\left| \alpha \right\rangle $ for $K=6,$ $\eta
=0.05, $ $|\xi |=100$ and different values of $|\alpha |$ as indicated. The
curve with $|\alpha |=6$ remains unchanged for $|\alpha |>6.$

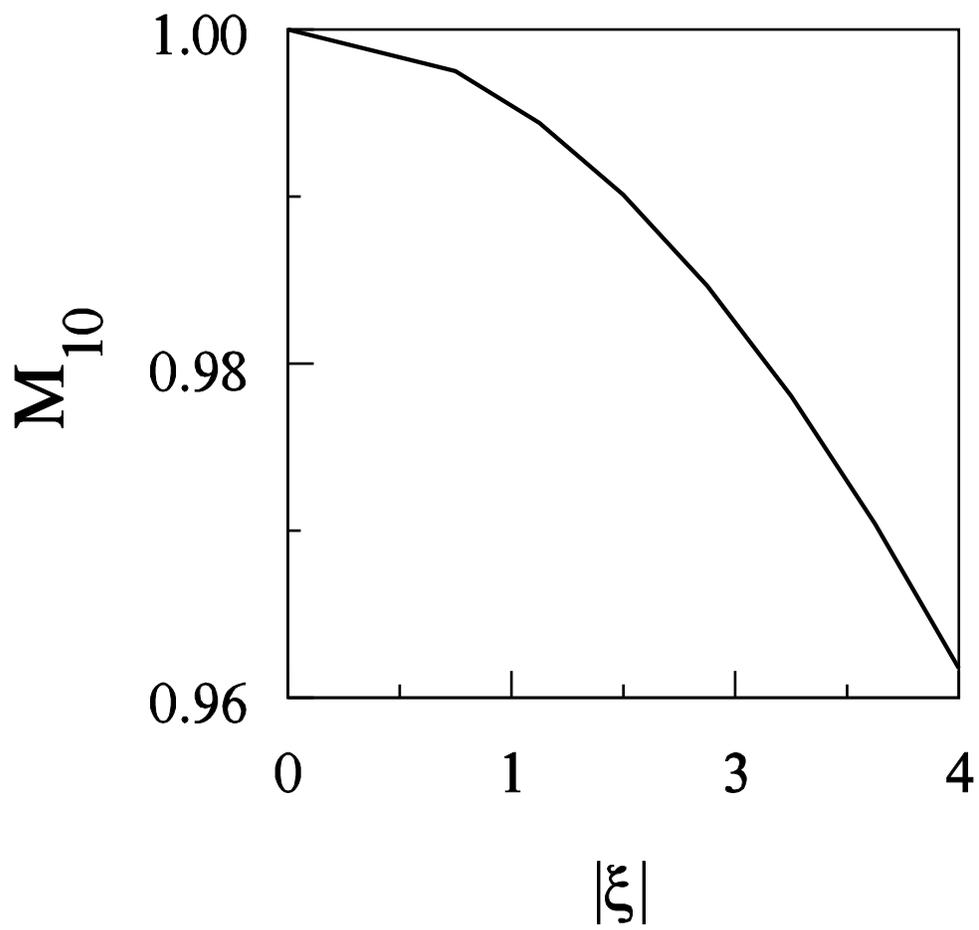
\item[Fig. 7:]  Mandel parameter against $|\xi |$ for $K=1,$ $j=0$ and $\eta
=0.05$ showing antibunching as opposed to the case with $f\equiv 1.$
Antibunching occurs as well for higher values of $\eta .$

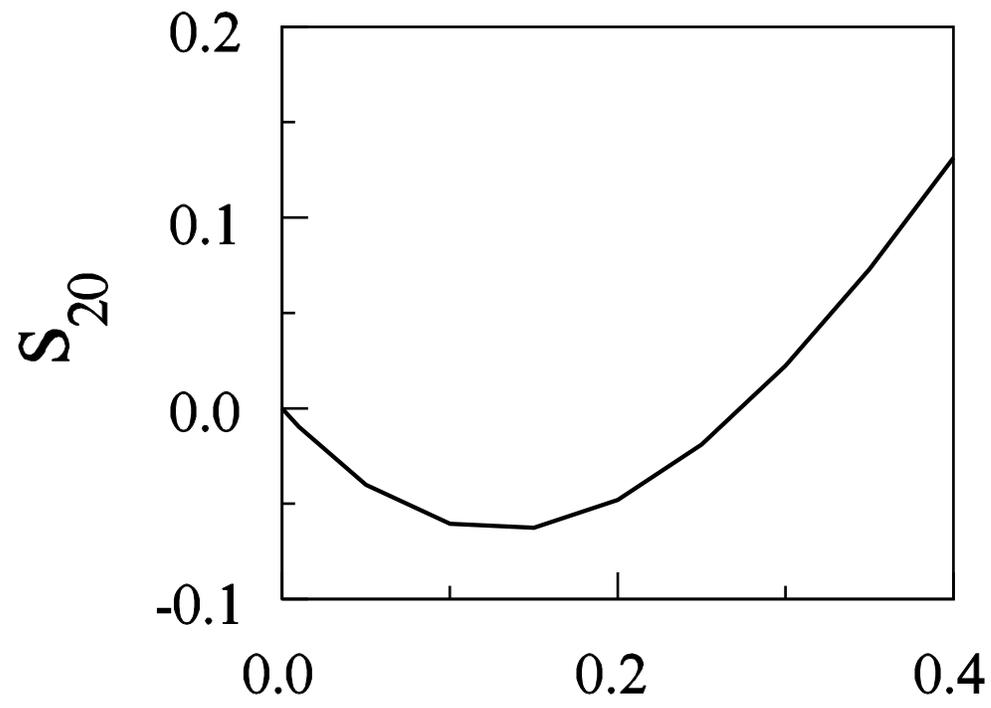
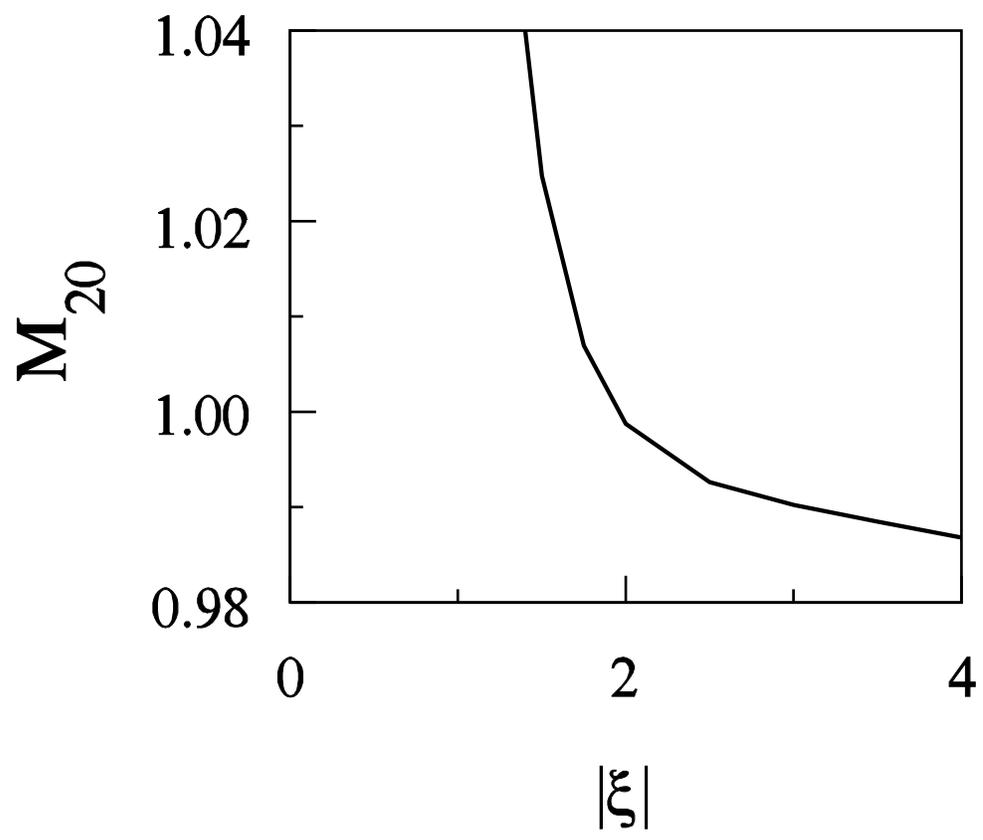
\item[Fig. 8:]  a) Squeezing to non-squeezing and b) super- to sub-poisson's
transition as $|\xi |$ increases for the state $\left| \xi
;20,f\right\rangle $ with $\eta =0.05$ (similarly for higher $\eta $).

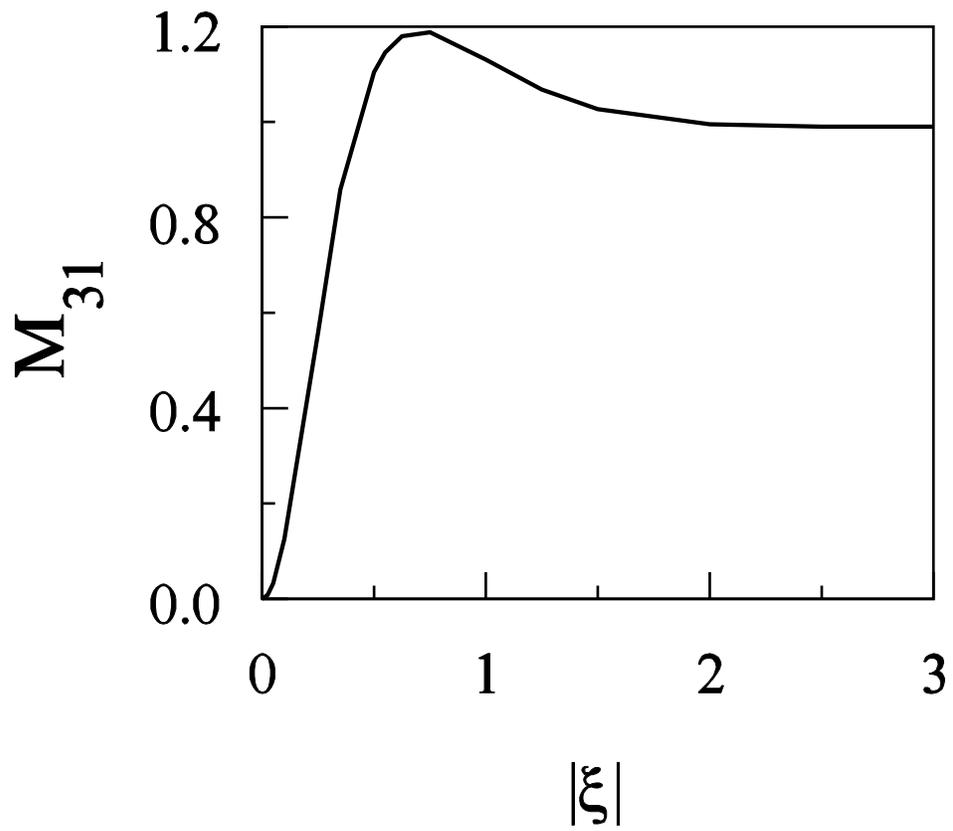
\item[Fig. 9:]  Mandel parameter as a function of $|\xi |$ for $K=3,$ $j=1$
and $\eta =0.05$ (similarly for higher $\eta $): the transition from
sub-poisson's to super-poisson's and back to sub-poisson's in the course of
increasing $|\xi |.$
\end{description}

\end{document}